\documentclass[conference]{IEEEtran}

\usepackage{amsthm}
\usepackage{amsmath}
\usepackage{amsfonts}
\usepackage[ruled,vlined]{algorithm2e}
\usepackage{algorithmic}
\usepackage{multicol}
\usepackage{multirow}
\usepackage{booktabs}
\usepackage{caption}
\usepackage{graphicx}
\usepackage{subfigure}
\usepackage{color}
\usepackage{bbm} 
\usepackage{enumerate}
\usepackage{url} 

\newcommand{\ie}{{\sl i.e.}}

\newcommand{\wrt}{{\sl w.r.t. }}
\newcommand{\aka}{{\sl a.k.a. }}

\begin{document}
\title{\textbf{DNA}: \textbf{D}ynamic Social \textbf{N}etwork \textbf{A}lignment}

%
\author{\IEEEauthorblockN{Li Sun, Zhongbao Zhang, Pengxin Ji, Jian Wen, Sen Su}
\IEEEauthorblockA{
\textit{Beijing University of Posts and Telecommunications, Beijing, China}\\
\{l.sun, zhongbaozb, jpx, j.wen, susen\}@bupt.edu.cn}
\and
\IEEEauthorblockN{Philip S. Yu}
\IEEEauthorblockA{
\textit{University of Illinois at Chicago, IL, USA}\\
psyu@uic.edu}
}

%

\maketitle

%
\begin{abstract}
  Social network alignment, aligning different social networks on their common users, is receiving dramatic attention from both academic and industry. 
  All existing studies consider the social network to be static and neglect its inherent dynamics. 
  In fact, the dynamics of social networks contain the discriminative pattern of an individual, which can be leveraged to facilitate social network alignment.
  Hence, we for the first time propose to study the problem of aligning dynamic social networks. 
  Towards this end, we propose a novel Dynamic social Network Alignment (DNA) framework, a unified optimization approach over deep neural architectures, to unfold the fruitful dynamics to perform alignment. 
  However, it faces tremendous challenges in both modeling and optimization: 
(1) To model the intra-network dynamics, 
  we explore the \emph{local dynamics} of the latent pattern in friending evolvement and the \emph{global consistency} of the representation similarity with neighbors.
 We design a novel deep neural architecture to obtain the \emph{dual embedding} capturing local dynamics and global consistency for each user. 
(2) To model the inter-network alignment, we exploit the underlying identity of an individual from the dual embedding in each dynamic social network. We design a unified optimization approach interplaying proposed deep neural architectures to construct a common subspace of \emph{identity embeddings}. 
(3) To address this optimization problem, we design an effective alternating algorithm with solid theoretical guarantees.
  We conduct extensive experiments on real-world datasets and show that the proposed DNA framework substantially outperforms the state-of-the-art methods.
\end{abstract}


%

%

\section{Introduction}
Social network alignment is to align different social networks on their common users, which is also known as anchor link prediction \cite{kong2013inferring}. 
It routinely finds itself in a wide spectrum of applications, such as network fusion, link prediction, information diffusion and cross-domain recommendation. 
On the whole,  social network alignment paves the way for the broad learning across social networks. Thus, it is receiving increasing attention from both academic and industry.

\begin{figure}
\centering
    \includegraphics[totalheight=1.81in]{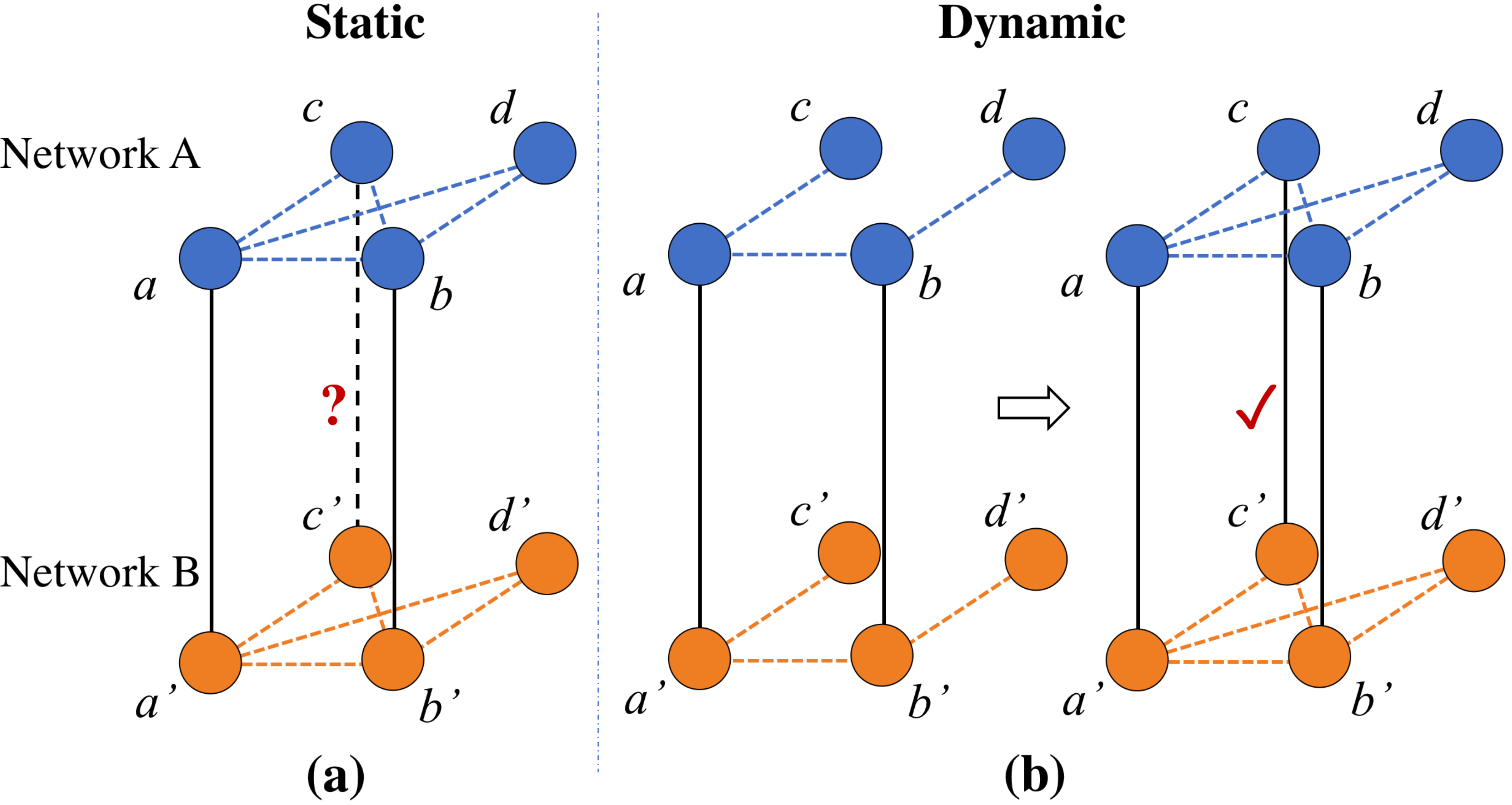}
      \vspace{-0.05in}
    \caption{Difference between static and dynamic social network alignment. Nodes denote users. Dotted lines denote  intra-network links and full lines denote anchor links. Fig. (a) shows static social network alignment, where some anchor (common) users, \ie, $\{a,a'\}$ and $\{b,b'\}$, are known in advance, while Fig. (b)  shows dynamic social network alignment with social networks evolving in the real-world scenario.}
    \vspace{-0.2in}
\end{figure}

All existing studies \cite{zafarani2013connecting,tan2014mapping,liu2016aligning,man2016predict,zhou2018deeplink,li2018distribution,zhang2015cosnet,zhang2015UMA,Cao2016BASS,Sun2018master,Zhong2018CoLink,zhang2018mego2vec,wang2018user,xie2018unsupervised,zhou2019tansl,wang2019learning}, to the best of our knowledge, consider social networks to be static.
As shown in Fig. 1 (a), given static social networks with some anchor (common) users, it is impossible to distinguish the counterpart of $c$ in social network A from $c'$ and $d'$ in network B.
Indeed, static social network alignment is still far away from the real-world scenario, where there is nothing permanent except change. Social networks are inherently dynamic as an individual more often than not adds or deletes his or her friends. 
Hence, \emph{can we leverage the dynamics of social networks to facilitate alignment?} The answer is \emph{yes!} 
We take advantage of the order of friending in real-world dynamic social networks. 
The order matters! In fact, friendship evolvement of an individual across social networks is dominantly deterministic rather than stochastic, as evidenced by the study of social psychology \cite{Utz2012It}.
Recall the toy example in Fig. 1 (b). 
We observe that $c$ first friends anchor user $\{a,a'\}$ and then friends $\{b,b'\}$ in network A, and $c'$ also friends  $\{a,a'\}$ prior to $\{b,b'\}$ in network B. 
It shows that $c$ and $c'$ share the same discriminative pattern of dynamics. 
Thus, $c$ is more likely to be aligned with $c'$ than $d'$.  
Motivated by this observation, we for the first time pose \emph{the problem of aligning dynamic social networks}.

Towards this end, we propose a novel Dynamic social Network Alignment framework, referred to as DNA, to unfold the fruitful dynamics of social networks to perform alignment. 
However, it faces tremendous challenges in both modeling and optimization:
\begin{itemize}
\item The first and foremost challenge lies in how to model the intra-network dynamics.
An individual, more often than not, expands or shrinks his or her friend list along with the evolvement	 of dynamic social networks.
However, no existing studies focus on the representing users with such complex dynamics for alignment. 
Thus, it is challenging to encode the intra-network dynamics. 

\item The second challenge is how to model the inter-network alignment.
An intuitive idea is to discover the underlying identity representation of users from different social networks to facilitate alignment.
Revealing such identity representation  is already nontrivial even if these social networks are simplified to be static \cite{liu2016aligning,man2016predict,mu2016user,Sun2018master,zhou2018deeplink}.
Thus, it is tougher to reveal the underlying identity from dynamic social networks.

\item Last but not the least, after modeling, it is always a challenge to address the optimization problem.
The optimization problem tends to be nonconvex due to the inherent complexity of dynamic network alignment.
For a nonconvex optimization, it lacks an effective method in general to approach the optimal or suboptimal solution.
Thus, it renders this problem much more challenging.
\end{itemize}

To address the first challenge, we design a novel deep neural architecture to model users in each dynamic social network.
To model users with dynamics, we exploit not only the temporal evolvement in friending behaviors of a user,  referred to as \emph{local dynamics}, but also representation similarity with the neighbors of the user, referred to as \emph{global consistency}. 
Specifically, to capture local dynamics, we design an LSTM Autoencoder zooming in the temporal friendship evolvement of each user for encoding the evolvement pattern. 
To capture global consistency, we design a consistency regularization and impose it onto the heart of the LSTM Autoencoder for compensating local dynamics. 
We utilize the hidden state in the heart of the neural architecture, referred to as \emph{dual embedding}, to jointly embed local dynamics and global consistency for the users of dynamic social network.



To address the second challenge, we design a unified optimization approach to construct a common subspace from different dynamic social networks.
All behaviors of an individual across different social networks tend to share something in common, termed as underlying identity.
It can be leveraged to identify this individual. 
Motivated by this fact, we first introduce a projection matrix to transform the dual embedding above into the \emph{identity embedding} for the underlying identity representation. 
Then, we formulate a unified optimization approach to construct the common subspace of identity embeddings, where users are naturally aligned according to the underlying identity.
This optimization is constrained on the collection of anchor users known in advance so that identity embeddings of the same individual remain consistent.
Finally, we obtain the unified optimization approach interplaying proposed deep neural architectures, referred to as \emph{the DNA framework}, for dynamic social network alignment.

The optimization problem of DNA is not jointly convex over the matrix variables of  (1) neural architecture parameters for dual embedding, (2) projection and (3) identity embedding.
To address this nonconvex optimization, we design an effective alternating algorithm to approach a local optimum of the  objective, \ie, we update one of the matrix variables while fixing others. 
Specifically, for the matrices of neural architecture parameters, we employ the gradient method.
For the matrices of projections, we derive the closed-form solution and prove that the derived solution is the global optimum.
For the matrices of identity embeddings, we first construct an auxiliary function and then derive the multiplicative updating rule. We further prove the correctness and convergence of the multiplicative updating rule and  draw corresponding theorems with solid theoretical guarantees.





In DNA, the proposed deep neural architecture enjoys the powerful nonlinear learning ability to capture the complex dynamics, and 
the unified optimization takes advantage of its inherent merit to naturally model the underlying identities.
The proposed DNA framework interplays two neural architectures with the unified optimization, thereby effectively aligning dynamic social networks.

To evaluate DNA, we first restore the real-world dynamic social network by framing a sequence of network snapshots via the collected timestamps. Then, we conduct extensive experiments on real-world datasets and demonstrate the superiority of the proposed DNA framework. 

Finally, we summarize noteworthy contributions as follows:


\begin{itemize}
\item To the best of our knowledge, we for the first time pose the problem of aligning dynamic social networks.
\item To address this problem, we design a novel DNA framework. In DNA, we propose a novel deep neural architecture to capture the complex dynamics and a unified optimization approach to construct the common subspace. 
\item To address the optimization of DNA, we design an effective alternating algorithm to approach an optimum with solid theoretical guarantees. 
\item We conduct extensive experiments on real-world datasets and experimental results demonstrate evident superiority of the proposed DNA framework.
\end{itemize}



The rest of this paper is organized as follows: in Sec. 2, we formalize this problem. Sec. 3 and Sec. 4 introduce the modeling and optimization of the proposed DNA framework, respectively. Sec. 5 shows the experimental results and Sec. 6 summarizes the related work. Finally, we conclude our work in Sec. 7.

\section{Problem Definition}

There are a source dynamic social network and a target dynamic social network. 
While the terminology of social network is commonly used, we use social graph in the rest of this paper to avoid ambiguity with the terminology of neural network. 
A dynamic social graph is naturally represented as a graph tensor. 
We use the superscript of $s$ (or $t$) to indicate notations of the source graph tensor $\mathcal G^{s}$ (or target graph tensor $\mathcal G^{t}$). 
Take the social graph tensor $\mathcal G^{s} \in \mathbb R^{N^{s} \times N^{s} \times M}$ for instance. 
There are $N^{s}$ users $u_{(\cdot)}^{s}$ in total and $M$ snapshots in time order. 
A slide of the tensor $\mathcal G^{s,m} \in \mathbb R^{N^{s} \times N^{s} }$ represents the $m^{th}$ snapshot of the dynamic social graph. 
The matrix $\mathcal G^{s,m}$ can be  asymmetric or symmetric 
and its element $[\mathcal G^{s,m}]_{ij}$ indicates directed or undirected connection strength between the $i^{th}$ user and $j^{th}$ user at the time of $m^{th}$ snapshotting in  $\mathcal G^{s}$. 
The value of $[\mathcal G^{s,m}]_{ij}$ can be either binary $\{0,1\}$ or arbitrary nonnegative value. 
The proposed framework, DNA, can align any (un)directed (un)weighted dynamic social graphs indeed.
Without loss of generality, the source and target dynamic social graphs are partially overlapped via the anchor users,
who join in both social graphs.
These users are recorded in an anchor set $\mathcal A=\{(k,l)_a\}$. Each element in $\mathcal A$ represents an anchor user, \ie, the $a^{th}$ element  $(k, l)_a$ describes the $k^{th}$ user in the source graph $\mathcal G^s$ corresponds to the $l^{th}$ user in the target graph $\mathcal G^t$.
We use $A=|\{(k,l)_a\}|$ to denote the size of $\mathcal A$.

We summarize the main notations in Table 1 and formally define the problem of aligning dynamic social networks:
\indent \newtheorem*{DNA}{Problem Definition (Aligning Dynamic Social Networks)} 
\begin{DNA}
Given the source and target dynamic social graph tensors $ { \mathcal G^{s}} $   and $ { \mathcal G^{t}} $ of $M$ snapshots with an anchor set $\mathcal A $, the problem of aligning dynamic social networks is to find a mapping $\Phi$ which maps a user to its owner of natural individual, \ie, $\Phi(u^{s}_{(\cdot)})=\Phi(u^{t}_{(\cdot)})$ holds if and only if the individual joins in both dynamic social graphs.
\end{DNA}
\vspace{-0.05in}
\begin{center}
  \begin{table}
    \centering
    \label{notation}
        \caption{Main Notations and Definitions}
    \begin{tabular}{c|l}
         \toprule
      Notations & Definition \\
       \midrule
      ${ \mathcal G}$  & the dynamic social graph tensor \\
      $\mathcal  A$ & the anchor set with size of $A$\\
      $\mathbf U$ & the matrix of dual embedding\\
      $\mathbf V$ & the matrix of identity embedding\\
      $\mathbf P$ & the indication matrix\\
      $\mathbf Q$ & the projection matrix\\
       $D^u$ & the dimension of dual embedding \\
      $D^c$ & the dimension of identity embedding \\
      $\alpha, \beta, \gamma$ &the model parameters\\
     \bottomrule
    \end{tabular} 
      \vspace{-0.15in}
  \end{table}
\end{center}
     \vspace{-0.3in}

\section{DNA: Framework}
To address the problem of aligning dynamic social networks, we propose a novel \textbf{D}ynamic social \textbf{N}etwork \textbf{A}lignment framework, referred to as \textbf{DNA}, with a unified optimization approach over deep neural architectures illustrated in Fig. 2. 
In the DNA framework, to encode the intra-network dynamics, we propose a novel deep neural architecture to embed users of each dynamic social network.
To address the inter-network alignment, we design a unified optimization framework interplaying proposed neural architectures to construct a common subspace, where user embeddings of the same individual are naturally aligned.

\subsection{Modeling Users with Complex Dynamics}




To model the complex dynamics, we explore the  temporal evolvement of the user's friendship, \ie, \emph{local dynamics}.
Moreover, user should own similar embedding with his or her friends to facilitate alignment as pointed out in studies \cite{zhang2015cosnet,man2016predict}.
That is, the \emph{global consistency}, representation similarity with the user's close neighbors, is of significant as well.

How can we capture both of them? 
Recently, LSTMs have proved to be effective for encoding evolvement in the sequential data \cite{GravesJ14}. 
However, it largely remains open to model users with complex dynamics for social network alignment. 
To bridge this gap, we design a novel deep neural architecture, LSTM Autoencoder with consistency regularization,
where we can obtain the dual embedding $\mathbf u^{(\cdot)}_i$ of dimension $D^u$ for each user 
jointly embedding local dynamics and global consistency.
Since $\mathcal G^{s}$ is the same as $\mathcal G^{t}$ on encoding the dynamics, we use $\mathcal G^{s}$ for elaboration in this subsection.

\begin{figure*}
\centering
    \includegraphics[totalheight=1.71in]{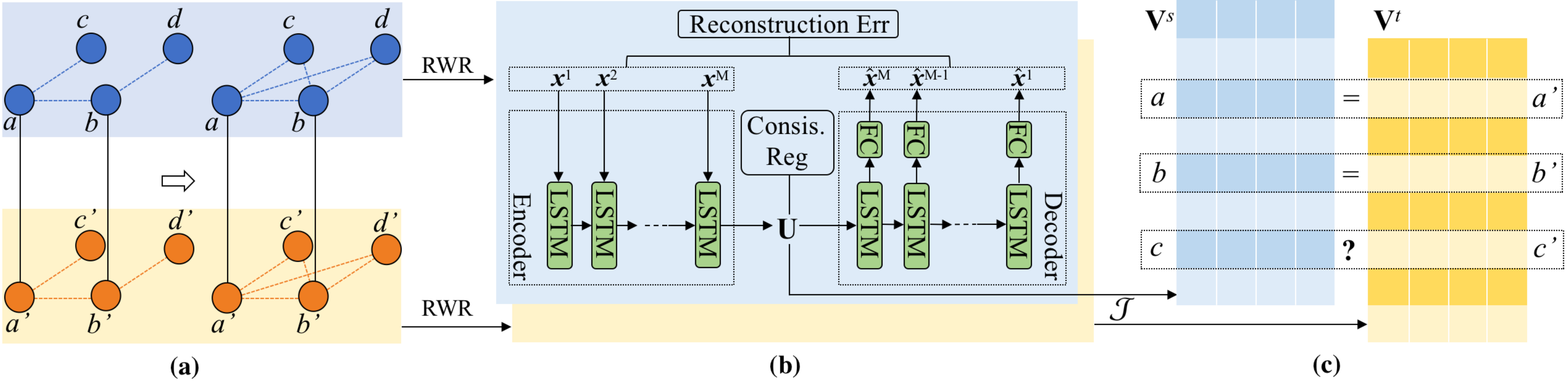}
    \vspace{-0.03in}
    \caption{
   An overview of DNA framework. The input are source and target dynamic social graphs with a collection of common users known in advance, as shown in Fig. (a). For each dynamic graph, ego vector sequences are fed into the proposed deep neural architecture to encode users with complex dynamics, as shown in Fig. (b), where superscript $s$ or $t$ is omitted for clarity. Finally, interplaying the proposed deep neural architectures, DNA framework outputs a common subspace of identity embeddings for alignment, as shown in Fig. (c).
    }
    \vspace{-0.2in}
\end{figure*}

\subsubsection{Ego Networks}




What should be the input for capturing the dynamics? 
We introduce the concept of an \emph{ego network} for each user and explore the temporal proximity evolvement in his or her ego network via an LSTM Autoencoder. 

Random walk with restart (RWR) \cite{tong2006fast} has been shown as an effective strategy to identify the proximities. 
Here, we leverage $\omega$-step RWR to calculate the proximity between users. 
Given an adjacency matrix $\mathbf G \in \mathbb R^{N^s \times N^s}$ and a start node $i$, indicated by a binary $\mathbf r_{i}^{(0)} \in \mathbb R^{N^s}$,
the RWR score vector $\mathbf r_{i} \in \mathbb R^{N^s}$ encodes the proximities between user $i$ and the others. 
Formally, $\mathbf r_{i}$ is defined as  $\mathbf r_{i}=\sum\nolimits_{\omega} \mathbf r_{i}^{(\omega)}$ 
and
\vspace{-0.03in}
\begin{equation}
\mathbf r_{i}^{(\omega)}=\xi \mathbf r_{i}^{(\omega-1)}\mathbf D^{-1} {\mathbf G}+ (1-\xi)\mathbf r_{i}^{(0)},
\label{rwr}
\vspace{-0.05in}
\end{equation}
\noindent where $\mathbf D=\text{diag}(\{d_i\})$ with $d_i=\sum\nolimits_j [{\mathbf G}]_{ij} $   and   $\xi$ is the probability that the random walker will not restart. 

Specifically, for the $i^{th}$ user, 
we first choose $W$ friends with high proximity to form the ego network.
That is, we locate the indexes $\{q_1, q_2, \dots, q_W\}$ of $W$ largest RWR scores on the aggregated adjacency matrix  ${\mathcal G}^{s, A}= \sum\nolimits_m e^{m-M}{\mathcal G}^{s, m}$ with exponential time decay for matrix $\mathbf G$ in Eq. (1). 
Now, the ego network is formed by connecting each node to its $W$ largest RWR score neighbors over the aggregated adjacency matrix.  

Next, we examine how the RWR scores to these $W$ friends evolve over the $M$ snapshots.  
Specifically, we calculate the ego vector $\mathbf x_i^{s,m} \in \mathbb R^{W}$ of user $i$ for each snapshot ${\mathcal G}^{s,m}$.
The ego vector $\mathbf x_i^{s,m}$ contains the proximity between user $i$ and his (or her) $W$ friends in the ego network, 
\ie, ${[\mathbf x_i^{s,m}]}_w={[\mathbf r_i]}_{q_w}$ and $\mathbf r_i$ is the RWR score vector on snapshot ${\mathcal G}^{s,m}$,
where ${\mathcal G}^{s,m}$ serves as matrix $\mathbf G$ in Eq. (1).
Finally, we obtain the ego vector sequence $\{\mathbf x_i^{s,m}\}_{m=1}^M$ depicting the evolvement regarding friendship distribution in the ego network for the given user $i$.  
The ego vector sequence $\{\mathbf x_i^{s,m}\}_{m=1}^M$ is to feed into the encoder LSTM.


\subsubsection{Capturing Evolvement Pattern}
How can we capture the evolvement pattern, \ie, \emph{local dynamics}? 
For the $i^{th}$ user, 
when ego vector sequence $\{\mathbf x_i^{s,m}\}$ is processed by the \emph{encoder LSTM} Cell recursively from $1$ to $M$, 
the pattern of temporal evolvement of his or her friendship is captured in the hidden state $\mathbf u_{i}^{s,M} \in \mathbb R^{D^u}$,
where we have the function $\mathbf u_{i}^{s,m}=\text{LSTMCell}({\mathbf u_{i}^{s,m-1}},{\mathbf x_i^{s,m}})$ performed recursively and $\mathbf u_{i}^{s,0}$ is initiated to zero vector. 
In LSTM, gating mechanisms (forget gate ${\mathbf f}$, input gate ${\mathbf i}$  and output gate ${\mathbf o}$)  with cell state  ${\mathbf c}$ are leveraged for long- and short-term dependence.
Formally, we describe transformations in the LSTM cell as follows:
\vspace{-0.03in}
\begin{equation}
\begin{aligned}
\mathbf f_{i}^{s,m} &= \sigma ( {\mathbf W}_f  [   \mathbf u_{i}^{s,m - 1}  ,  \mathbf x_{i}^{s,m} ] +  \mathbf b_f ), \\
\mathbf i_{i}^{s,m}  &= \sigma ({\mathbf W}_i   [    \mathbf u_{i}^{s,m - 1}  ,  \mathbf x_{i}^{s,m}  ] + \mathbf b_i),\\
\mathbf o_{i}^{s,m} &= \sigma ({\mathbf W}_o  [    \mathbf u_{i}^{s,m - 1}  ,   \mathbf x_{i}^{s,m}] + \mathbf b_o),\\
\tilde{\mathbf c}_{i}^{s,m} & = \text{ReLU} (   {\mathbf W}_c    [ \mathbf u_{i}^{s,m - 1}   ,  \mathbf x_{i}^{s,m}   ] +    \mathbf b_c),\\
\mathbf c_{i}^{s,m}   &=  \mathbf f_{i}^{s,m}   \odot  \mathbf c_{i}^{s,m - 1} +    \mathbf i_{i}^{s,m} \odot \tilde{\mathbf c}_{i}^{s,m},\\
\mathbf u_{i}^{s,m}  & = \mathbf o_{i}^{s,m}    \odot \text{ReLU} ( \mathbf c_{i}^{s,m}  ),
\end{aligned}
\vspace{-0.07in}
\end{equation}
\noindent where $[\cdot, \cdot]$ denotes concatenation. 
$\mathbf W_{(\cdot)}$ and $\mathbf b_{(\cdot)}$ are parameter matrices and biases, respectively. 
$\sigma$ is the sigmoid function and $ \odot$ is the Hadamard product. 
We utilize the activation function $\text{ReLU}$ to facilitate further optimization. 
The \emph{decoder LSTM} decodes the hidden state $\mathbf x_i^{s,M}$ to recover ego vector reversely from $M$ to $1$, and we append a full connection layer to the decoder for dimension transformation, as shown in Fig. 2 (b).

For all users in the dynamic social graph $\mathcal G^s$, we give the reconstructed error of LSTM Autoencoder in the form of tensor approximation: 
\begin{equation}
\vspace{-0.06in}
\left\| \mathcal X^{s}-\hat{\mathcal X}^{s}\right\|_F^2,
\label{recons}
\vspace{-0.03in}
\end{equation}
where $||\cdot||_F$ is the Frobenius norm. $\mathcal X^{s} \in \mathbb R^{N^{s} \times W \times M}$ is the ego tensor. Its $m^{th}$ slide $\mathcal X^{s,m} \in \mathbb R^{N^{s} \times W}$ is the matrix of ego vector for the $m^{th}$ snapshot $\mathcal G^{s,m}$, whose $i^{th}$ row is the ego vector $\mathbf x_{i}^{s,m}$. $\hat{\mathcal X}^{s}$ is the reconstructed ego tensor from decoder LSTM.
The evolvement pattern is embedded in the output hidden state of the last encoder LSTM cell $\mathbf u_{i}^{s,M}$. 
According to the study \cite{wang2016structural}, this embedding space will preserve the local dynamics in the graph tensor
via minimizing the reconstructed error even without an explicit preservation restriction.



\subsubsection{Consistency Regularization}
Focusing on the temporal evolvement of the user friendship, 
aforementioned embedding $\mathbf u^{s}$ highlights local dynamics but lacks \emph{global consistency}.
As a compensation, we impose a consistency regularization $\mathcal R^{s}$ onto the heart of LSTM Autoencoder as shown in Fig. 2 (b), to incorporate the global consistency.
Specifically, we draw neighbors closer in the embedding space via a pairwise penalty $\mathcal R^{s}_{ij} = r_{ij}\left\|\mathbf u_i^{s}- \mathbf u_j^{s}\right\|_2^2$,
where $r_{ij}=[\mathbf r_i]_j$ indicates the proximity between the $i^{th}$ user and $j^{th}$ user in $\mathcal G^{s}$.
The consistency regularization $\mathcal R^{s}$ is summation of $\mathcal R^{s}_{ij}$ over all user pairs $\{(ij)\}$ in $\mathcal G^{s}$,
\begin{equation}
\vspace{-0.03in}
\mathcal R^{s}= Tr\left[{\mathbf U^{s}}^T {\mathcal G^{s,L}} \mathbf U^{s}\right],
\label{reg}
\vspace{-0.03in}
\end{equation}
where $\mathcal G^{s,L}=\mathbf D-\mathcal G^{s,A}$ is the graph Laplacian and $\mathbf U^{s} \in \mathbb R_+ ^{N^{s} \times D^{u}}$ is the \emph{dual embedding} matrix of local dynamics and global consistency, whose $i^{th}$ row corresponds to the $i^{th}$ user.

For dynamic social graph $\mathcal G^{s}$, combining Eq. (\ref{recons}) and Eq. (\ref{reg}), we obtain the loss function of the proposed deep neural architecture as follows:
\begin{equation}
\mathcal D^{s}=\left\|\mathcal X^{s}-\hat{\mathcal X}^{s}\right\|_F^2+\alpha \mathcal R^{s},
\label{lar}
\vspace{-0.03in}
\end{equation}
where $\alpha$ is the weight parameter. 
The norm term is for local dynamics and the trace term is for global consistency.

\subsection{Constructing Common Subspace}


All behaviors of an individual in both social graphs tend to share a common nature, his or her underlying identity. It can be leveraged to identify this individual. 
Motivated by this fact, we design a unified optimization approach to construct the common subspace of  \emph{identity embedding} $\mathbf v_i^{s}\in \mathbb R_+^{D^c}$ with dimension $D^c$ depicting such underlying identity. 

 
Specifically, we first introduce a projection matrix $\mathbf Q^{(\cdot)} \in \mathbb R^{D^c \times D^u}$, for each dynamic social graph, to transform the dual embedding $\mathbf u_i^{s}\in \mathbb R_+^{D^u}$ into the identity embedding $\mathbf v_i^{s}$. 
Use source dynamic graph $\mathcal G^s$ for elaboration. Formally, we have $\mathbf u_i^{s}=\mathbf v_i^{s}\mathbf Q^{s}$ for the $i^{th}$ users.
Thus, for $\mathcal G^{s}$, we obtain the matrix of identity embedding $\mathbf V^{s} \in \mathbb R_+ ^{N^s \times D^c}$ ($D^c \ll \min\{N^{s},N^{t}\}$),
whose $i^{th}$ row is the identity embedding $\mathbf v_i^{s}$ for user $i$. 
We formulate this transformation as a loss term in the form of matrix approximation: 
\begin{equation}
\left\|\mathbf U^{s} - \mathbf V^{s}\mathbf Q^{s}\right\|_F^2.
\label{uv}
\end{equation}
Note that, $D^c$, without loss of generality, is not necessarily equal to $D^u$ for dual embedding.




Then, we construct the common subspace of the identity embeddings.
Naturally, the identity embeddings of the same individual in different networks remain equal, \ie, the matrices of identity embedding  $\mathbf V^{s}$ and $\mathbf V^{t}$ are aligned on the anchor users.
To formulate the constraint, we derive a pair of indication matrices $\{\mathbf P^{s}, \mathbf P^{t} \}$ from the anchor set $\mathcal A=\{(k,l)_a\}$, where $\mathbf P^{s} \in \mathbb R^{A \times N^{s}}$, $ \mathbf P^{t} \in \mathbb R^{A \times N^{t}}$ and $A$ is the size of the anchor set $\mathcal A$. 
The indication matrix
$ \mathbf P^{s} $ is defined as $\left[\ \mathbf p^s_1, \ \mathbf p^s_2, \dots ,\ \mathbf p^s_A \ \right]^T$
whose row vector $\mathbf p^s_{a} \in \mathbb R^{N^{s}}$ is 
a binary vector with only one element valued $1$, \ie, ${[\mathbf p^s_a]}_k=1$. The vector $\mathbf p^s_a$ is to indicate the corresponding row, the identity embedding $\mathbf v_k^s$ of $k^{th}$ user in $\mathcal G^s$, for aligning.
Similarly, $ \mathbf P^{t} $ is defined as $\left[\ \mathbf p^t_1, \ \mathbf p^t_2, \dots ,\ \mathbf p^t_A \ \right]^T$
and ${[\mathbf p^t_a]}_l=1$ is the only element valued $1$ in binary vector $\mathbf p^t_a \in \mathbb R^{N^{t}}$.
We enforce the following equality
\begin{equation}
\mathbf P^{s}\mathbf V^{s}=\mathbf P^{t}\mathbf V^{t},
\end{equation}
for all anchor users in $\mathcal A$ to depict the alignment between matrices of identity embeddings $\mathbf V^{s}$ and $\mathbf V^{t}$.

\subsection{Optimization Objective}
Now, we formulate a unified optimization interplaying the proposed deep neural architectures for dynamic social network alignment. 
Incorporating Eq. (\ref{lar}) and Eq. (\ref{uv}) for each dynamic social graph $\mathcal G^{(\cdot)}$, we have following objectives:
\begin{equation}
\begin{aligned}
\mathcal J^{s} &=   {\mathcal D^{s}} +  \beta ||\mathbf U^{s} - \mathbf V^{s}\mathbf Q^{s}||_F^2,\\
\mathcal J^{t}& =   {\mathcal D^{t}} +  \beta ||\mathbf U^{t} - \mathbf V^{t}\mathbf Q^{t}||_F^2,
\end{aligned}
\end{equation}
where $\beta$ weighs the significance of the projection. 

Next, we rewrite the equality constraint in the form of matrix approximation and reformulate the equality constraint as a penalty term with a coefficient $\gamma$ sufficiently large.

Finally, we obtain the unified optimization problem of DNA framework, which interplays two deep neural architectures for dynamic social network alignment, as follows:
\begin{equation}
\mathop {\min }\limits_{\mathbf V^{(\cdot)},\mathbf Q^{(\cdot)}, \mathbf \Theta^{(\cdot)} } \ \ {\mathcal J} = {\mathcal J^{s}}+{\mathcal J^{t}}+\gamma  \left\| \mathbf P^{s}\mathbf V^{s} - \mathbf  P^{t}\mathbf V^{t}\right\|_F^2,
\label{J}
\end{equation}
where $\mathbf \Theta^{(\cdot)}$ denotes parameter matrices of the proposed neural architecture.
The users whose identity embeddings are positioned closely in the common subspace are regarded as good candidates for the social network alignment.






\section{DNA: Optimization}
The optimization problem of DNA in Eq. (\ref{J}) is not jointly convex over matrix variables $\mathbf Q$, $\mathbf V$ and $\mathbf \Theta$ for both dynamic social graphs. Thus, it is infeasible to obtain the closed-form solution of the global optimum.
To bridge this gap, we design an effective alternating algorithm to approach an optimum. The core idea is to approach a (local) optimal solution  \wrt  one matrix variable while fixing others. $\mathbf \Theta$ is readily to be solved via gradient methods. Obviously, the key challenge lies in the updating rules for $\mathbf Q$ and $\mathbf V$. The derivation of updating rule of the variables associated with $\mathcal G^{t}$ is the same as that of $\mathcal G^{s}$ and thus omitted due to the limit of space. The overall procedure of the DNA framework is summarized in Algorithm 1. 


\subsection{Updating $\mathbf Q$}

To solve $\mathbf Q^{s} = \mathop {\arg }\limits_{\mathbf Q^{s}} \min \mathcal J({\mathbf Q^{s}})$, we give the closed-form solution as follows:
\begin{equation}
\mathbf Q^{s} =\left({\mathbf V^{s}}^T \mathbf V^{s}\right)^{-1}{\mathbf V^{s}}^T\mathbf U^{s}
\label{q}
\end{equation}

\newtheorem*{theorem0}{Theorem (Optimal Solution)}
\begin{theorem0}
Fixing other matrix variables, the updating rule in the proposed DNA for ${\mathbf Q^{s}}$ and ${\mathbf Q^{t}}$ gives the global optimal solution to minimize overall objective $\mathcal  J$ \wrt ${\mathbf Q^{s}}$ and ${\mathbf Q^{t}}$, receptively.
\end{theorem0}

\begin{proof}
\noindent First, we derive the matrix of partial differentials of $\mathcal J$ \wrt $\mathbf Q^{s}$ as follows:
\begin{equation}
\nabla_{\mathbf Q}^{s} =-2 {\mathbf V^{s}}^T \mathbf U^{s} +2{\mathbf V^{s}}^T \mathbf V^{s}\mathbf Q^{s}
\end{equation}
Then, we study the Hessian matrix
$\mathbf H_{\mathbf Q}^{s}$ of $\mathcal J$ \wrt $\mathbf Q^{s}$:
\begin{equation}
\mathbf H_{\mathbf Q}^{s}=\mathbf I_k \otimes{\mathbf V^{s}}^T \mathbf V^{s},
\end{equation}

\begin{algorithm}
          \caption{DNA Framework}
          \LinesNumbered
            \KwIn{Source dynamic social graph tensor: ${ \mathcal G^{s} }$ \\
                       $\quad \ \  \quad$ Target dynamic social graph tensor: ${ \mathcal G^{t} }$\\
                         $\quad \ \  \quad$ Anchor set: $\mathcal A $}
            \KwOut{Candidate lists for alignment between ${ \mathcal G^{s} }$ and ${ \mathcal G^{t} }$}
            Generate ego tensor $\mathcal X^{s}$ and $\mathcal X^{t}$ for dynamic social graph tensor $\mathcal G^{s}$ and $\mathcal G^{t}$, respectively\;
            Pretrain the proposed deep neural architecture for each ego tensor via $\mathcal J_d^{s}$ and $\mathcal J_d^{t}$\;
            Perform forward propagation to obtain ${\mathbf U^{s}}^{(0)}$ and ${\mathbf U^{t}}^{(0)}$\;
            Initialize ${\mathbf V^{s}}^{(0)}= {\mathbf U^{s}}^{(0)}$ and ${\mathbf V^{t}}^{(0)}= {\mathbf U^{t}}^{(0)}$\;
            \While{not converging} {
              \For{each dynamic social graph}{
                  Back-propagation to update $\mathbf \Theta^{s}$ (or $\mathbf \Theta^{t}$)\;
              }
              \For{each dynamic social graph}{
                Forward propagation to obtain $\mathbf U^{s}$ (or $\mathbf U^{t}$)\;
                Update $\mathbf Q^{s} =\left({\mathbf V^{s}}^T \mathbf V^{s}\right)^{-1}{\mathbf V^{s}}^T\mathbf U^{s}$ (the updating rule of  $\mathbf Q^{t}$ is of the same structure)\;
                Update $\mathbf V^{s}$ (or $\mathbf V^{t}$) via Eqs. (\ref{v1}) and (\ref{v2})\;
              }
          }
          Generate candidate lists for alignment via computing the distance between the rows in $\mathbf V^{s}$ and $\mathbf V^{t}$.
\end{algorithm}

\noindent where $\otimes$ denotes the Kronecker product. It is block diagonal,
\begin{equation}
[\mathbf H_{\mathbf Q}^{s}]_{[k][k]}={\mathbf V^{s}}^T \mathbf V^{s},
\end{equation}
where $\mathbf I_k$ is a $k$-dimensional identity matrix and $[\mathbf H_{\mathbf Q}^{s}]_{[k][k]}$ is the $(k,k)$-th block in matrix $\mathbf H_{\mathbf Q}^{s}$. Obviously, the block diagonal $\mathbf H_{\mathbf Q}^{s}$ is always positive semidefinite as ${\mathbf V^{s}}^T \mathbf V^{s}$ is positive semidefinite.
Thus, $\mathcal J$ is convex \wrt $\mathbf Q^{s}$.
Therefore, the global optimum is achieved by setting $\nabla_{\mathbf Q}^{s}=\mathbf 0$. Finally, we obtain the closed-form updating rule for ${\mathbf Q^{s}}$ in Eq. (\ref{q}) as the inverse matrix always exists for any positive semidefinite matrix ${\mathbf V^{s}}^T \mathbf V^{s}$. 
\end{proof}


\subsection{Updating $\mathbf V$}
To solve $\mathbf V^{s} = \mathop {\arg }\limits_{\mathbf V^{s}} \min \mathcal J({\mathbf V^{s}})$, we give the multiplicative updating rule as follows:
\begin{equation}
\mathbf V^{s}=  \mathbf V^{s} \odot \sqrt{
\frac{  \beta (\mathbf \Psi^{s}+{\mathbf V}^{s} \mathbf \Gamma^{s}) + \gamma  \mathbf \Lambda^{s}   } {
  \beta  (\mathbf \Upsilon ^{s}+ {\mathbf V}^{s} \mathbf \Phi ^{s})+\gamma  \mathbf \Pi^{s}}},
\label{v1}
\end{equation}
\noindent where
\begin{equation}
\begin{array}{*{20}{c}}
\mathbf \Psi^{s} = ({\mathbf U^{s}}{\mathbf Q^{s}}^T)^+ &\ \mathbf \Upsilon ^{s}=  ({\mathbf U^{s}}{\mathbf Q^{s}}^T)^- \\
\mathbf \Phi ^{s}= ({\mathbf Q^{s}}{\mathbf Q^{s}}^T)^+ &\ \ \mathbf  \Gamma^{s}= ({\mathbf Q^{s}}{\mathbf Q^{s}}^T)^- \\
\mathbf \Pi ^{s}= {\mathbf P^{s}}^T{\mathbf P^{s}}{\mathbf V^{s}}\ &\mathbf \Lambda ^{s}={\mathbf P^{s}}^T{\mathbf P^{t}}{\mathbf V^{t}} \\
\end{array}
\label{v2}
\end{equation}
\noindent Note that, $\odot$ is the Hadamard product. Both $\frac{(\cdot)}{(\cdot)}$ and $\sqrt{\cdot}$ are pairwise operators. Nonnegative operators $(\cdot)^+$ and $(\cdot)^-$  are defined as follows:
\begin{equation}
\resizebox{0.85\hsize}{!}{$
\begin{array}{*{20}{c}}
{ [\mathbf X^+]_{ij}= \frac{|[\mathbf X]_{ij}|+[\mathbf X]_{ij}}{2} }&{[\mathbf X^-]_{ij}=\frac{|[\mathbf X]_{ij}|-[\mathbf X]_{ij}}{2},}
\end{array}
$}
\end{equation}

\noindent where $| \cdot |$ is the absolute value of a scalar. 
Hence,  $\mathbf X^+$ and $\mathbf X^-$ are absolute value matrices of positive elements and negative elements of an arbitrary matrix $\mathbf X$, respectively. 

Next, we prove the correctness of the multiplicative updating rule in Eqs. (\ref{v1}) and (\ref{v2}), and give the theorem below:

\newtheorem*{theorem1}{Theorem (Correctness)}
\begin{theorem1}
The limit point of multiplicative updating rules of DNA for ${\mathbf V^{s}}$ (as well as ${\mathbf V^{t}}$) in Eqs. (\ref{v1}) and (\ref{v2})  satisfies the Karush-Kuhn-Tucker (KKT) condition.
\end{theorem1}
\begin{proof}
We study the objective function $\mathcal J$ in Eq. (\ref{J}) \wrt $\mathbf V^{s}$, referred to as  $\mathcal J({\mathbf V^{s}})$. Utilize the fact  $||\mathbf X||_F^2=tr(\mathbf X^T\mathbf X)$ for any real matrix $\mathbf X \in \mathbb R^{n \times m}$. 
After some algebraic operations and omitting constants, we obtain:
\begin{equation}
\begin{aligned}
\mathcal J({\mathbf V^{s}})= & -2 \beta tr(({\mathbf V^{s}}{\mathbf Q^{s}})^T{\mathbf U^{s}}) 
+ \beta tr(({\mathbf V^{s}}{\mathbf Q^{s}})^T{\mathbf V^{s}}{\mathbf Q^{s}}) \\
& -2 \gamma  tr(({\mathbf P^{t}}{\mathbf V^{t}})^T{\mathbf P^{s}}{\mathbf V^{s}}) 
+\gamma  tr(({\mathbf P^{s}}{\mathbf V^{s}})^T{\mathbf P^{s}}{\mathbf V^{s}}).
\end{aligned}
\label{objv}
\end{equation}
We introduce the matrix $\mathbf \Omega$ of the Lagrangian multipliers to enforce the nonnegative constraint on $\mathbf V^{s}$, leading to the Lagrangian function of $\mathcal J({\mathbf V^{s}})$,
\begin{equation}
\begin{aligned}
\mathcal L({\mathbf V^{s}})= \mathcal J({\mathbf V^{s}}) + tr(\mathbf \Omega\mathbf V^{s}).
\end{aligned}
\end{equation}
Following the KKT complementary slackness condition, we obtain the fixed point equation as follows, 
\begin{equation}
[\nabla_{\mathbf V}^{s}]_{ij}[{\mathbf V^{s}}]_{ij}=[\mathbf \Omega]_{ij}[{\mathbf V^{s}}]_{ij}=0,
\label{kkt1}
\end{equation}
\noindent where $\nabla_{\mathbf V}^{s}$ is the matrix of  partial differentials:
\begin{equation}
\begin{aligned}
\nabla_{\mathbf V}^{s}= & -2 \beta {\mathbf U^{s}}{\mathbf Q^{s}}^T+ 2\beta{\mathbf V^{s}}{\mathbf Q^{s}}{\mathbf Q^{s}}^T \\
&-2\gamma {\mathbf P^{t}}^T{\mathbf P^{t}}{\mathbf V^{t}}
 +2\gamma {\mathbf P^{s}}^T{\mathbf P^{s}}{\mathbf V^{s}}.
\end{aligned}
\end{equation}
The solution must satisfy Eq. (\ref{kkt1}) in the limit.
Given the multiplicative updating rule in  Eqs. (\ref{v1}) and (\ref{v2}), we have, in the limit, ${\mathbf V^{s}}^{(\infty)}={\mathbf V^{s}}^{(\tau+1)}={\mathbf V^{s}}^{(\tau)}=\mathbf V^{s}$, where the superscript $^{(\tau)}$ indicates the $\tau$-th iteration. Utilizing the fact $\mathbf X = \mathbf X^{+}+ \mathbf X^{-}$, we obtain that:
\begin{equation}
[\nabla_{\mathbf V}^{s}]_{ij}[{\mathbf V^{s}}]^2_{ij}=0.
\label{kkt2}
\end{equation}
The right hand side in both  Eq. (\ref{kkt2}) and Eq. (\ref{kkt1}) is $0$. The left hand side in both Eq. (\ref{kkt2})  and Eq. (\ref{kkt1}) is a multiplication of two factors and the first factor in both equations is identical. As $[{\mathbf V^{s}}]^2_{ij}=0$ if and only if $[{\mathbf V^{s}}]_{ij}=0$, we can claim that Eq. (\ref{kkt2}) holds if and only if Eq. (\ref{kkt1}) holds. Thus, we reach the Theorem (Correctness).
\end{proof}

\begin{center}
  \begin{table}
    \centering
      \caption{Statistics of the datasets}
    \label{Table 4}
    \begin{tabular}{c|c| c |c }
      \hline
     {Dataset }& \#(Nodes) &  \#(Links) &  \#(Anchor Users)  \\
     \hline
    Twitter & $5,167$ & $164,660$ & \multirow{2}{*}{$2,858$}  \\
      
    Foursquare& $5,240$ &  $\ \ 76,972 $ &  \\
    
      \hline
    \end{tabular} 
     \label{stat}
  \end{table}
\end{center}

\section{Experiment}
\subsection{Experimental Setup}
\noindent \textbf{Datasets}:
We enrich a benchmark dataset \cite{liu2016aligning,kong2013inferring,Sun2018master,zhou2018deeplink,zhang2015UMA,li2018distribution}, Twitter-Foursquare (TF), by collecting time information of friending. We restore the real-world dynamic social network by framing a sequence of network snapshots via the collected timestamps. The statistics are listed in Table \ref{stat}.

\noindent \textbf{Comparison Methods}:
To evaluate the performance of the proposed DNA framework, we chose several state-of-the-art methods for comparison, which are introduced as follows:
\begin{itemize}
\item \textbf{PALE} \cite{man2016predict}: It leverages the structure information and performs an embedding-matching framework to align users across social networks.
\item \textbf{IONE} \cite{liu2016aligning}: It utilizes a unified optimization framework for aligning social networks subject to hard or soft constraint on the anchor users via an embedding approach. 
\item \textbf{MASTER} \cite{Sun2018master}: It constructs a common subspace of multiple social networks via a semi-supervised optimization framework. Note that, it only considers structure information in this paper for fair comparison.
\item \textbf{DeepLink} \cite{zhou2018deeplink}: It introduces a neural framework, which leverages dual learning to facilitate common subspace construction to perform alignment.
\end{itemize}
\noindent \textbf{Parameter Settings}: In the proposed DNA framework, 
we set $\omega$ to $3$ for $\omega-$step RWR. We utilize the $l_2$ norm to regularize the neural architecture parameters for alleviating overfitting and  employ the dropout trick in the training, whose empirical value is set to $0.8$. 

\noindent \textbf{Evaluation Metric}:
We evaluated all the comparison methods in terms of the following metrics:
\begin{itemize}
\item \emph{Precision$@K$} is defined as $ \frac {1}{N_A} \sum_{i=1}^{N_A} \mathbbm{1}_i \{success@K\} $, where $\mathbbm{1}_i \{success@K\}=1$ if and only if the groundtruth exists in the candidate list of length $K$, otherwise $\mathbbm{1}_i \{success@K\}=0$. $N_A$ is the number of groundtruth anchor users.
\item MAP$@K$ is defined as $ \frac {1}{N_A} \sum_{i=1}^{N_A } \frac{1}{Rank_i}$ , where $Rank_i$ is the rank of the candidate hitting the ground truth in the candidate list of length $K$. Specially, we set $\frac{1}{Rank_i}=0$ for not hitting. $N_A$ is defined in \emph{Precision$@K$}. Note that, MAP highlights the rank of the hitting candidate in a nonlinear way.
\end{itemize}
Higher value of the metrics signifies better performance.

      
      
      
    

\begin{figure} 
\centering 
\subfigure[Precision@K under different $K$]{
\includegraphics[width=0.49\linewidth]{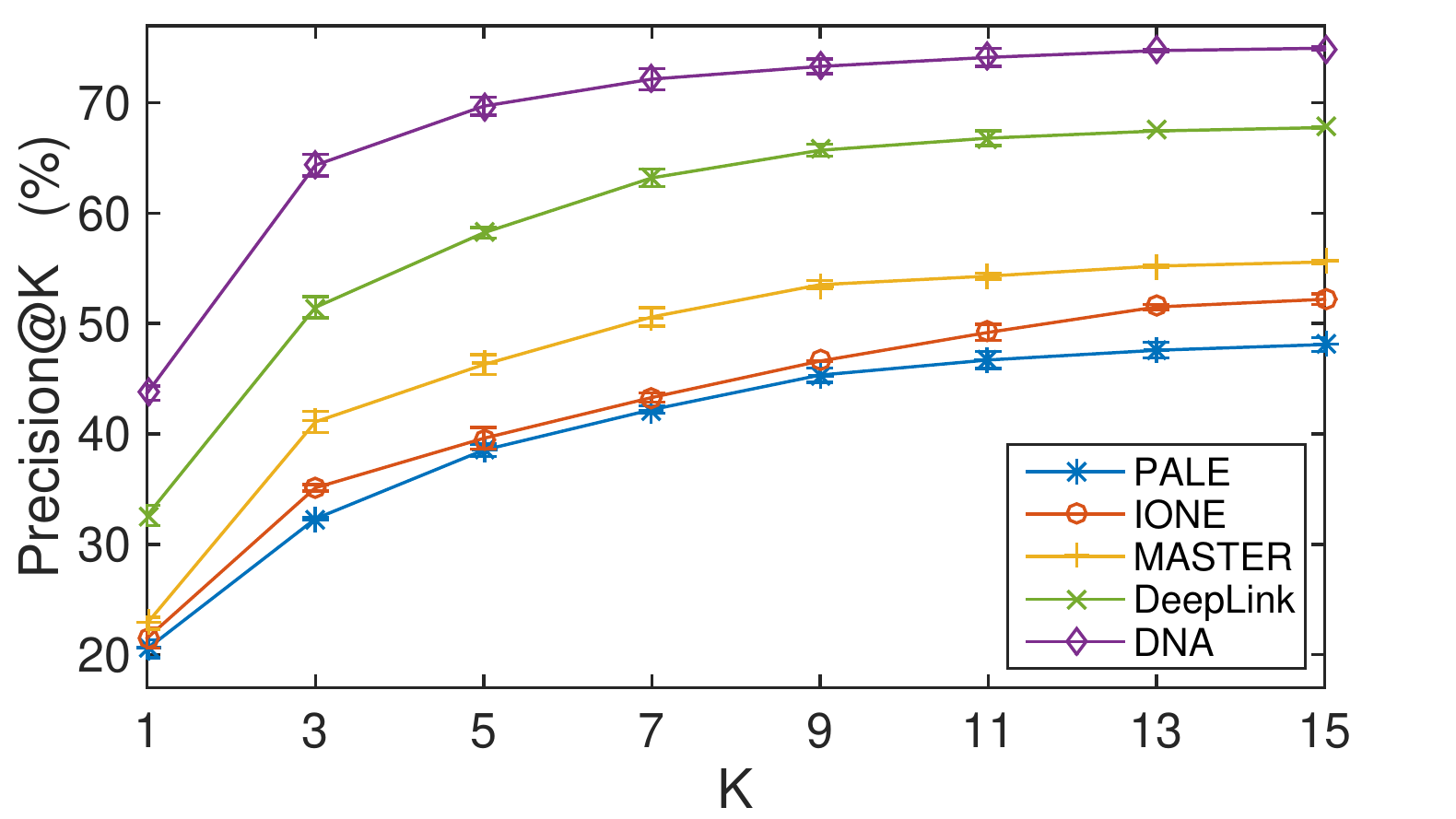}}
\hspace{-0.035\linewidth}
\subfigure[Precision@3 under different $\lambda$]{
\includegraphics[width=0.49\linewidth]{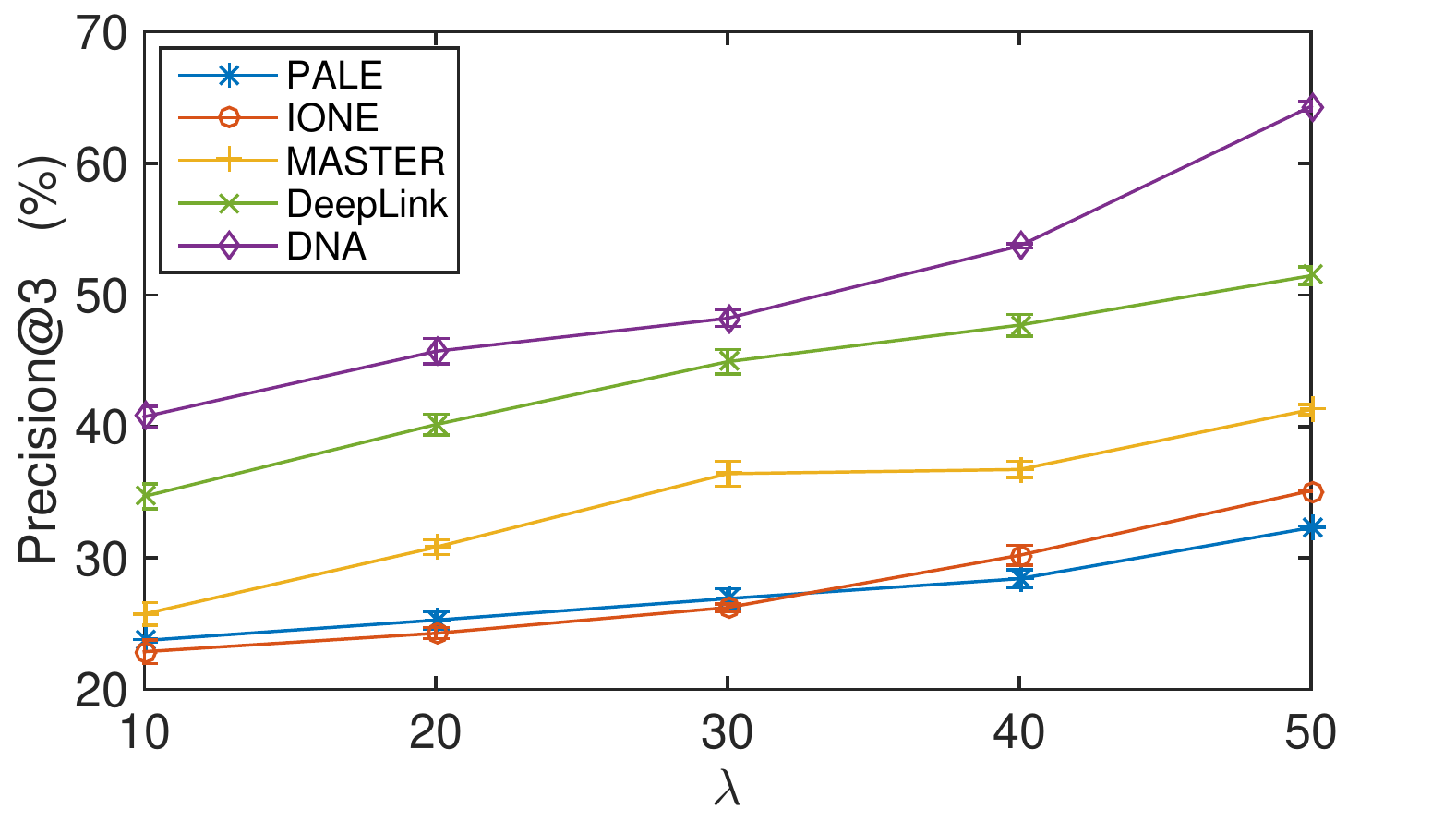}}
\vfill
\vspace{-0.05in}
\subfigure[MAP@K under different $K$]{
\includegraphics[width=0.49\linewidth]{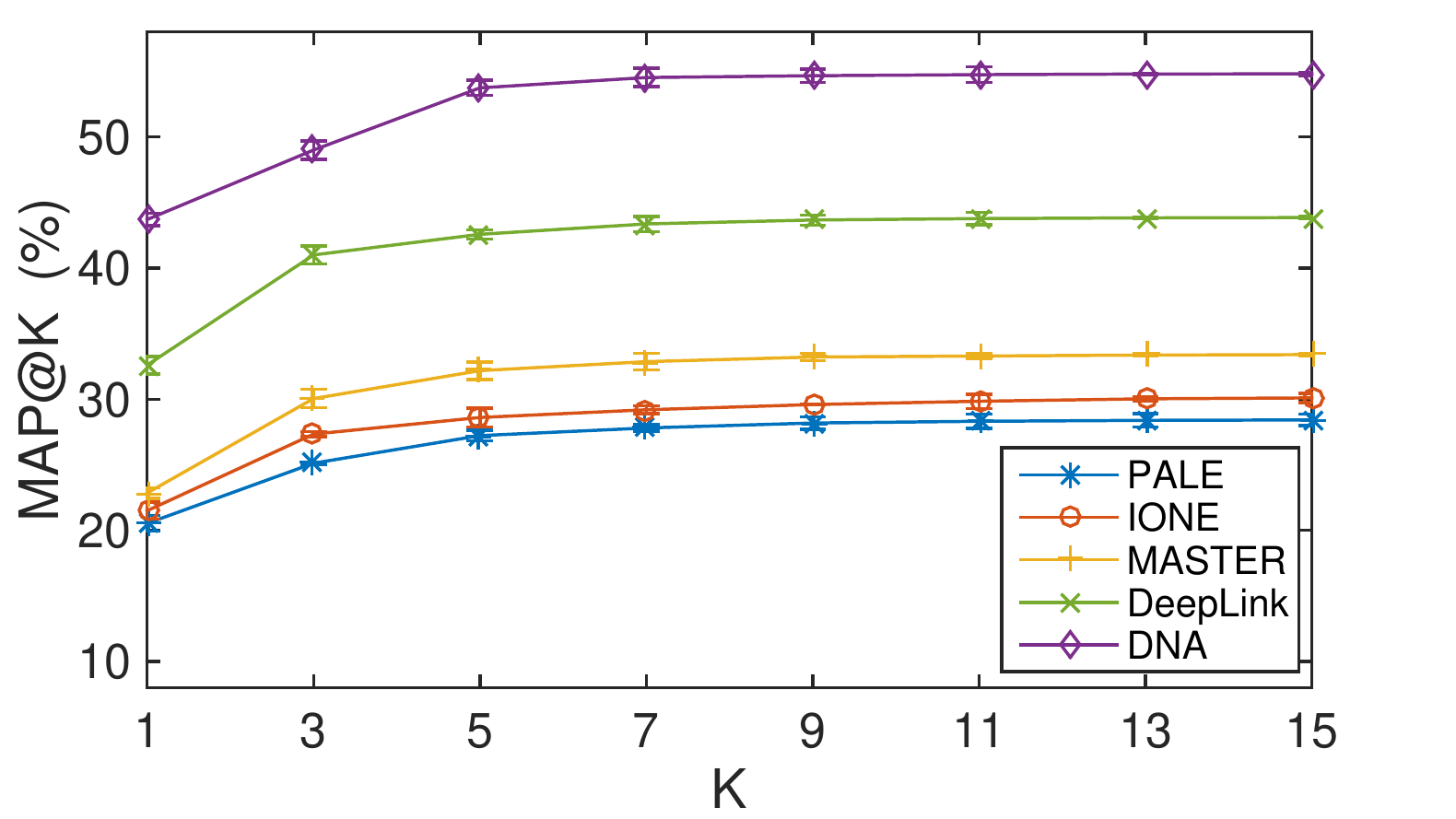}}
\hspace{-0.035\linewidth}
\subfigure[MAP@3 under different $\lambda$]{
\includegraphics[width=0.49\linewidth]{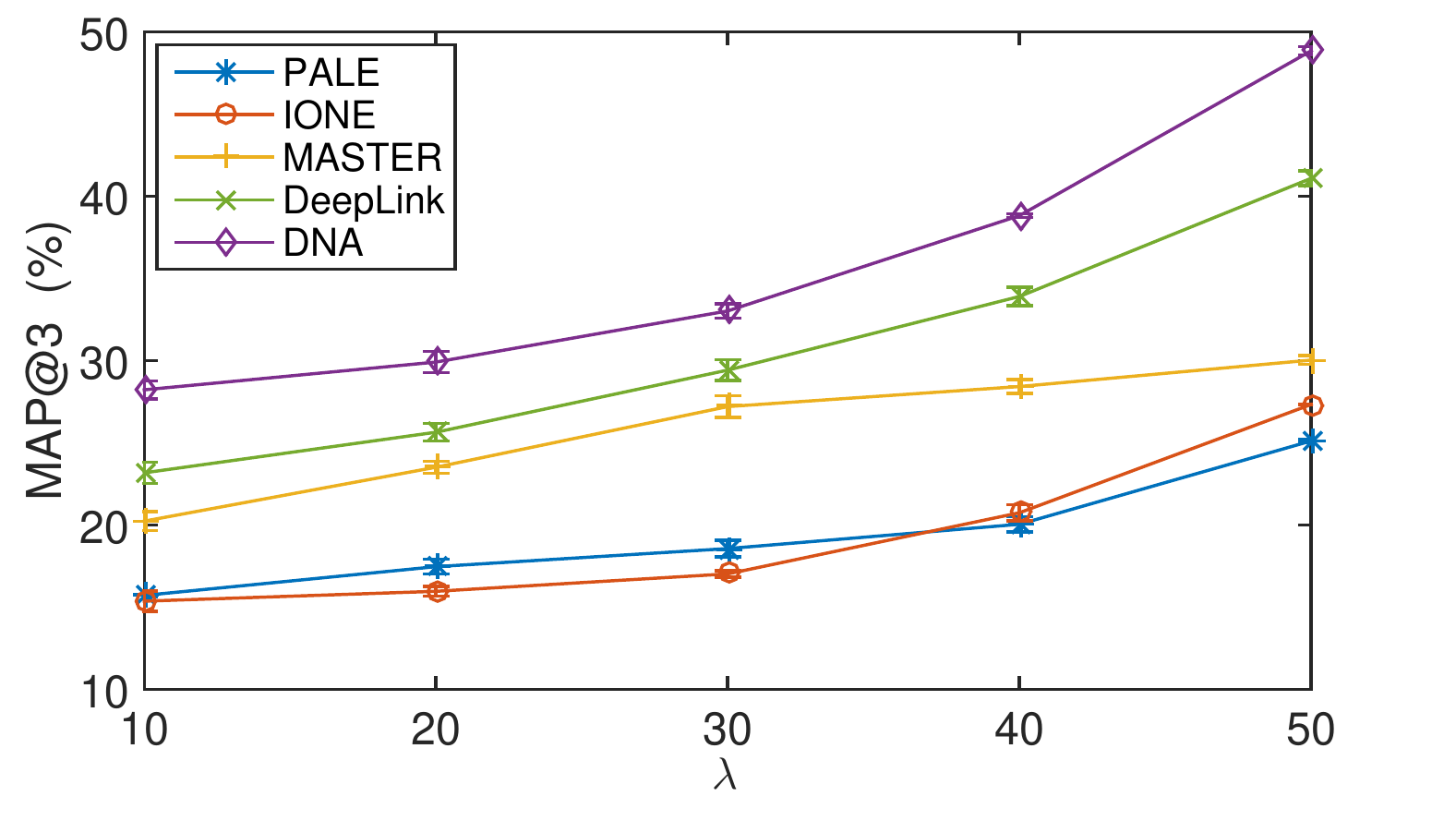}}
\vspace{-0.1in}
\caption{Experimental results on Twitter-Foursquare dataset}
\vspace{-0.25in}
\label{tf}
\end{figure}

\subsection{Experimental Results}
We first focus on verifying the importance of dynamics and its reason behind, and then discuss the supervision for DNA, where all the embedding dimensions are set to $128$ as default.  
Finally, we conduct sensitivity analysis on embedding dimension. 
We repeated each experiment 10 times and report the mean with $95\%$ confidence interval.  

\textbf{The dynamics is important!}
We compare the proposed DNA framework with static alignment methods to demonstrate the importance of dynamics.  
To achieve this goal, we evaluate DNA and its competitors under various settings. 
Specifically, on both datasets, 
we first vary the length $K$ of the candidate list from $1$ to $15$ on $50\%$-overlap datasets, reported in Fig. \ref{tf} (a) and (c).
We then vary the overlap rate $\lambda$ from $10$ to $50$ while fixing $K=3$, reported in Fig. \ref{tf} (d) and (d). 
The overlap rate $\lambda$ is measured by $\frac{2A}{S+T}$, where $A$, $S$ and $T$ are the number of anchor users, source network users and target network users, respectively. A $\lambda$-overlap dataset is generated by randomly deleting users. 
The performance in terms of both precision and MAP are evaluated.

Obviously, DNA consistently outperforms all comparison methods.
For instance, DNA achieves $1.96 \times$ precision and $1.94 \times$ MAP of PALE on the $50\%$-overlap benchmark (TF) dataset.
The reasons are two-fold: 
(1) The DNA framework captures the complex dynamics in the real-world social network evolvement, which is discriminative for social network alignment.  
However, the other methods cannot effectively identify the alignment due to limited user depiction, which inevitably loses the fruitful information residing in network dynamics.
(2) DNA constructs a common subspace modeling the underlying identities, and the users are thereby naturally aligned in the common subspace.
However, other methods, such as IONE \cite{liu2016aligning}, PALE \cite{man2016predict} and DeepLink \cite{zhou2018deeplink}, are somehow powerless to reveal underlying identities of individuals for alignment.
To sum up, experimental results verify that dynamics significantly facilitates social network alignment. 


\begin{figure} 
\centering 
\subfigure[Precision@3]{
\includegraphics[width=0.49\linewidth]{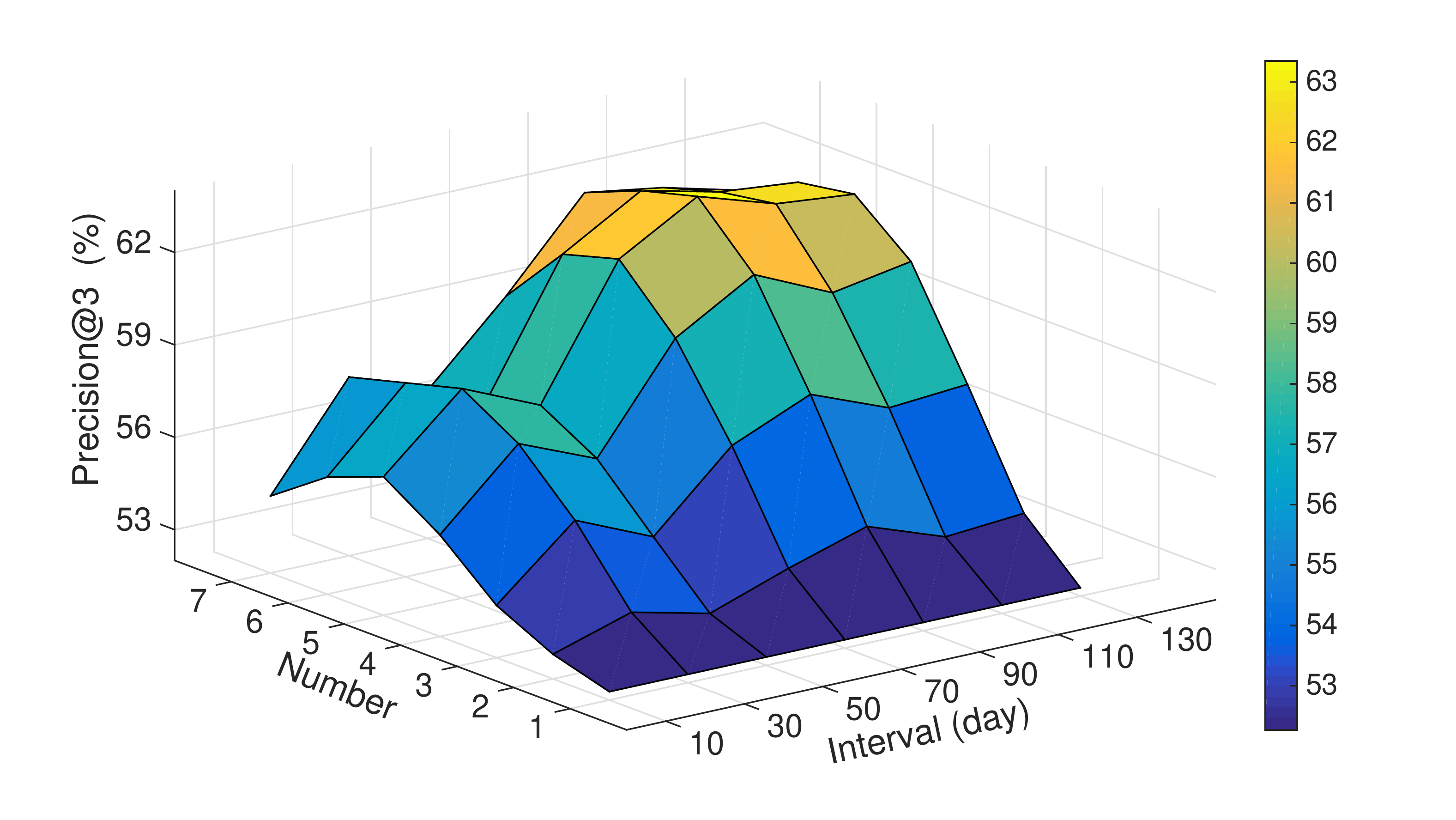}}
\hspace{-0.035\linewidth}
\subfigure[Precision@3]{
\includegraphics[width=0.49\linewidth]{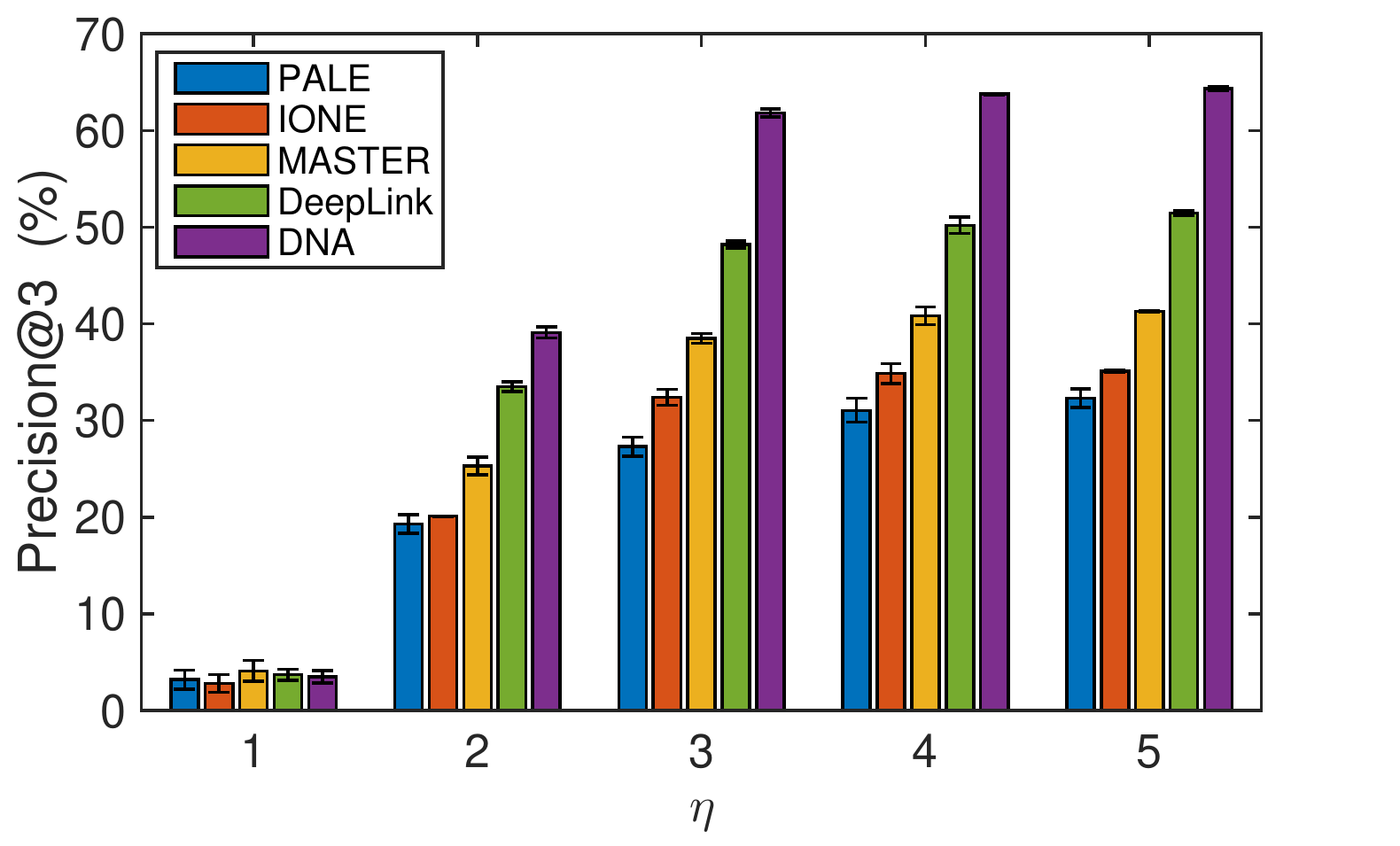}}
\vfill
\vspace{-0.05in}
\subfigure[MAP@3]{
\includegraphics[width=0.49\linewidth]{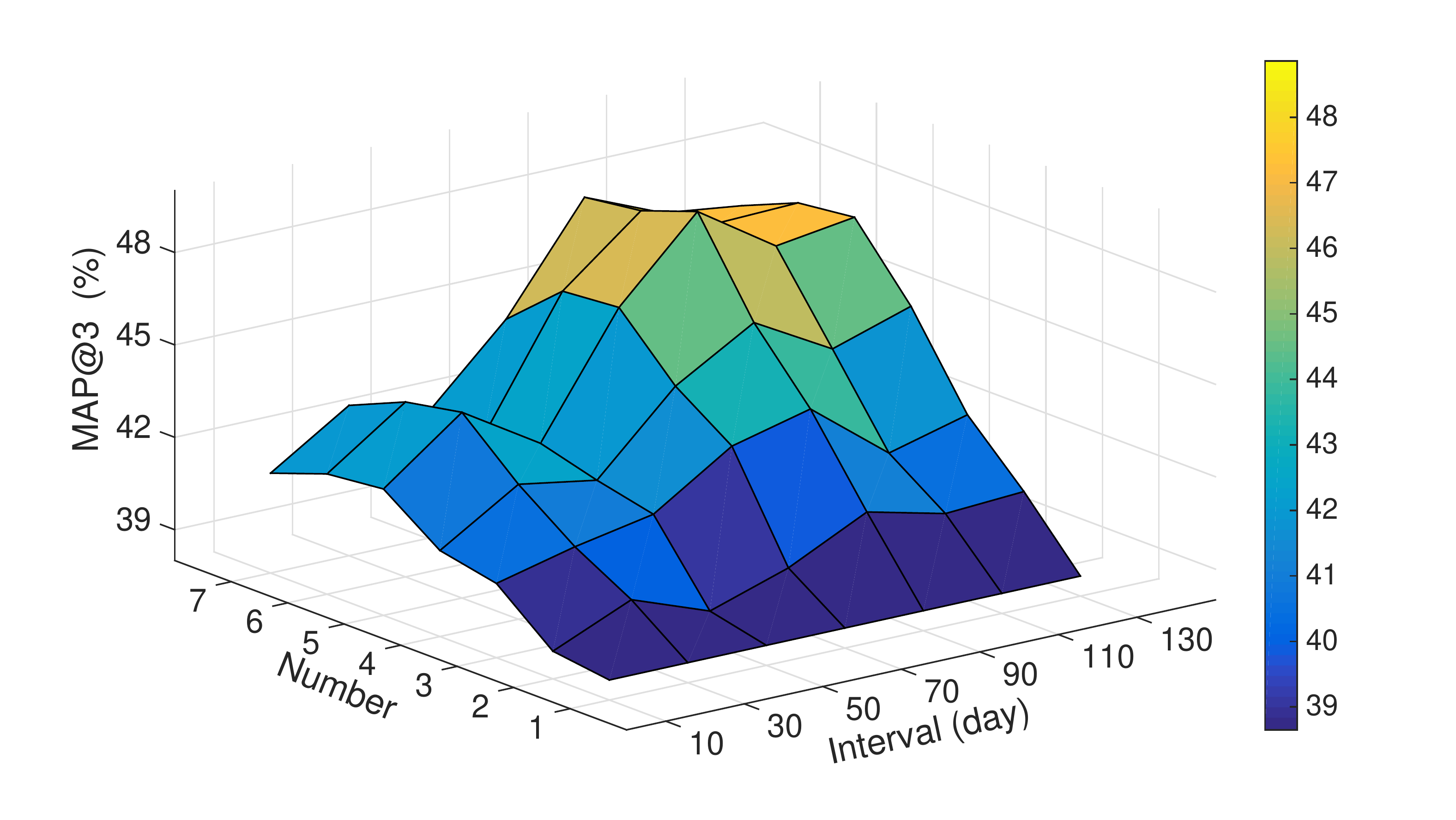}}
\hspace{-0.035\linewidth}
\subfigure[MAP@3]{
\includegraphics[width=0.49\linewidth]{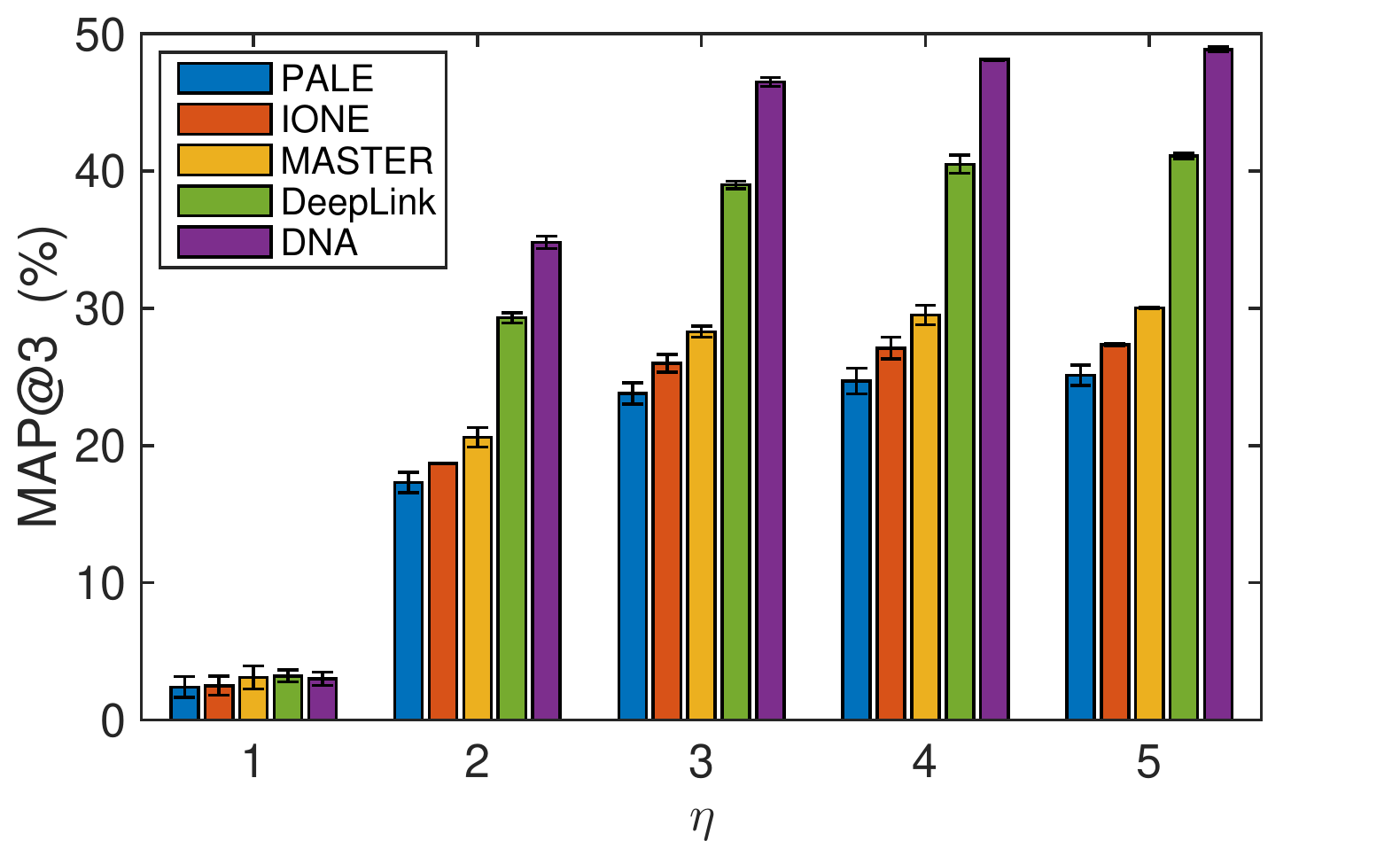}}
\vspace{-0.1in}
\caption{Parameter sensitivity on Twitter-Foursquare dataset}
\vspace{-0.25in}
\label{para}
\end{figure}

\textbf{On the Snapshotting:}
The performance of DNA leverages  $5$ snapshots with the interval of $3$ months in Fig. \ref{tf}. Now, we zoom in the dynamics of  social networks by varying the frequency and number of snapshots. 
A natural question arises that: the more the snapshots, the better the performance? 
We evaluate the performance of DNA under various settings of snapshotting, reported in Fig. \ref{para} (a) and (c), where the confidence interval is omitted for clarity. 
Both Precision$@3$ and MAP$@3$ of DNA increases as frequency and number of snapshots increase in general. 
However, the performance tends to be saturated when the information collected is adequate.
It is reasonable indeed. 
In general, the higher frequency and higher number of snapshots better restore the network dynamics, \ie, friending behaviors in this context. 
It is of high discriminative ability for alignment as it is dominantly correlated to the patterns of individuals acting across social networks as evidenced in the study \cite{Utz2012It}.
For instance, if an individual more often than not expands or shrinks his or her friend list once a few weeks, it will reveal this pattern in a few months.
Thus, we need a time window long enough and a snapshotting frequency high enough to capture the pattern of friending or collaboration behavior.
Moreover, we observed that outdated information may not further help analysis as stated in the study \cite{Utz2012It} as well.
These results in fact verify the motivation of our work and in turn shed light on the dynamics of  behavior pattern for the further study of social psychology.

\begin{figure} 
\centering 
\subfigure[Precision@3]{
\includegraphics[width=0.49\linewidth]{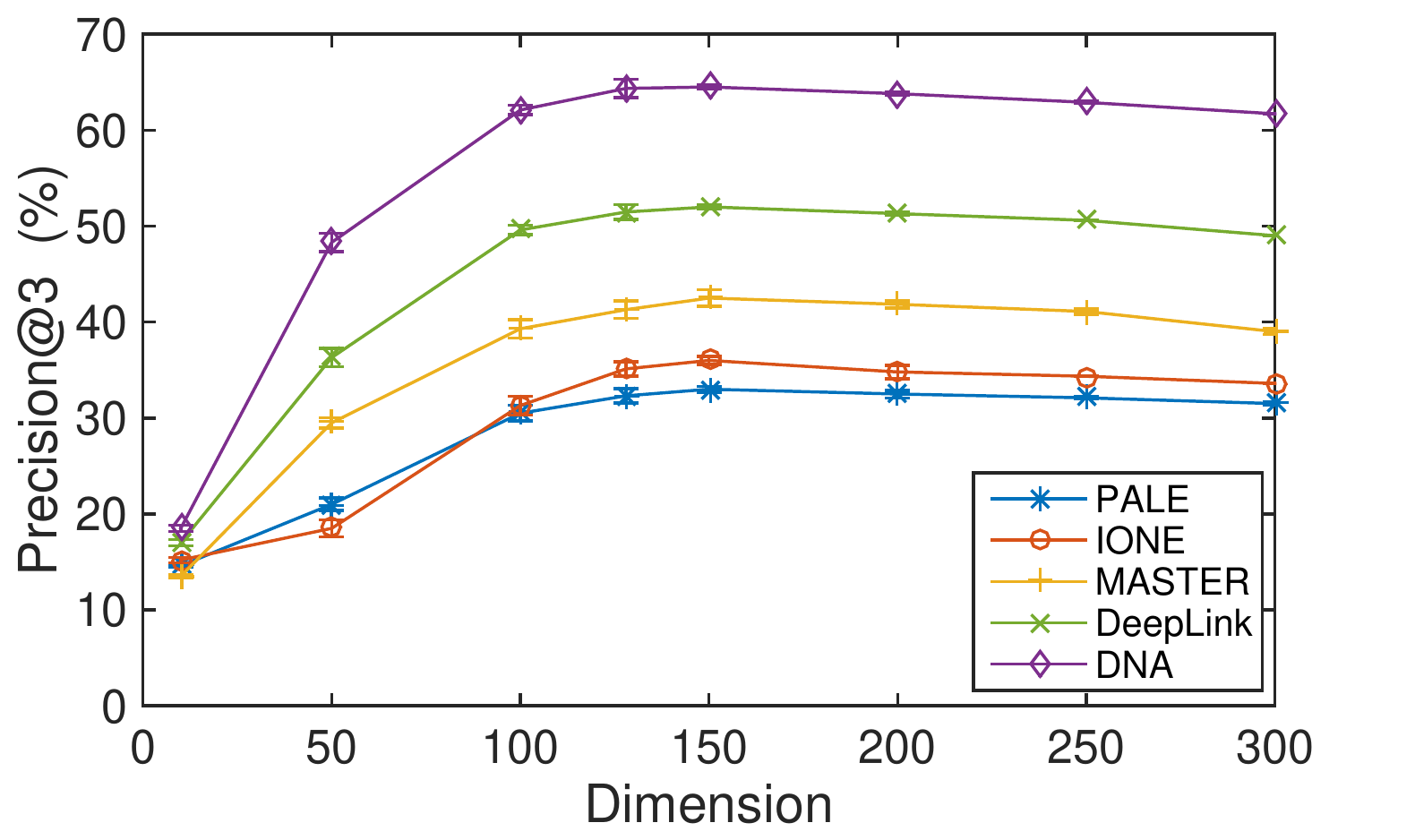}}
\hspace{-0.035\linewidth}
\subfigure[MAP@3]{
\includegraphics[width=0.49\linewidth]{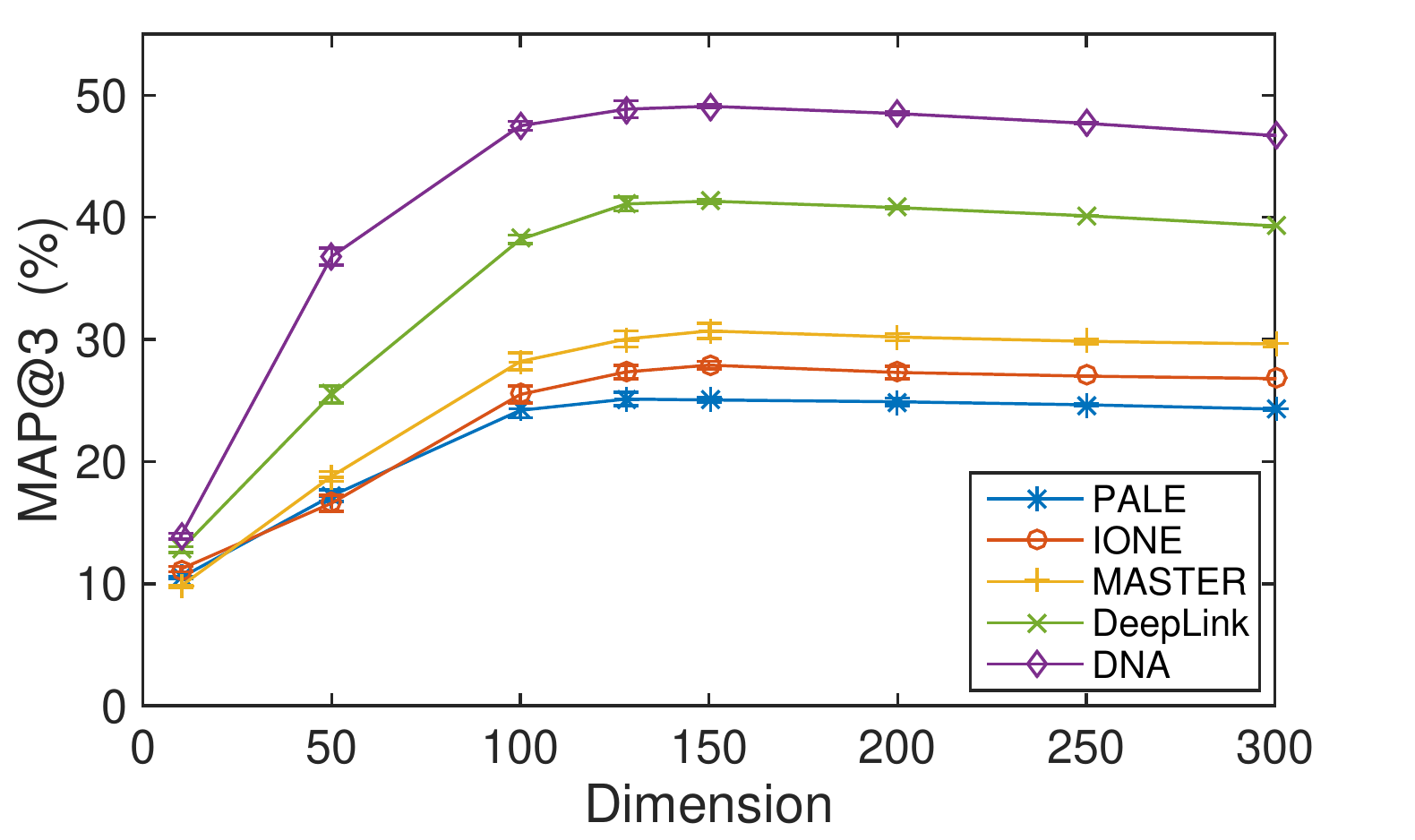}}
\caption{Performance under different embedding dimensions}
\label{d}
\vspace{-0.25in}
\end{figure}

\textbf{On the Supervision Information:} 
We analyze the impact of training rate $\eta$ by varying $\eta$ from $1\%$ to $5\%$.
As shown in Fig. \ref{para} (b) and (d), the performance of DNA rises dramatically as $\eta$ increases from $1\%$ to $3\%$, and then saturates when $\eta$ exceeds $3\%$. 
On both TF and AN datasets with $50\%$ overlap rate, DNA achieves better performance in terms of precision and MAP than its competitors, leveraging only a few anchor users.
It is expected and the reason lies in two folds:
 (1) In the DNA framework, the user dynamics are encoded in a self-supervised way, \ie, we can obtain the embedding space without supervision. 
 (2) The common subspace will be constructed by aligning the embedding spaces on only a few anchor users, which is one of the inherent merits of DNA. 
 To summarize, the experimental results on different training rates verify the learning ability of the proposed DNA framework.


\textbf{On the Embedding Dimension:}
We further discuss the impact of the embedding dimension. The higher the embedding dimension, the better the performance? It is not always the case.
We vary the dimension from $10$ to $300$ and report the precision and MAP on both TF and AN datasets in Fig. \ref{d}.
As shown in these subfigures, both Precision$@3$ and MAP$@3$ of all these methods are undesirable when the dimension is low, and tends to be better as the dimension increases. 
The reason lies in that when user embeddings with low dimensionality tend to entangle densely, thus resulting in the mislead of alignment. 
However, we will not obtain further performance gain when embedding dimension is large enough to layout user embeddings in the common embedding space.
Moreover, the DNA framework consistently outperforms all comparison methods as the generated embeddings encode the discriminative pattern residing in the dynamic social networks, which significantly facilitates  social network alignment.

\vspace{0.01in}
\section{Related Work}
We summarize the related works in the following two fields:

\subsection{Social Network Alignment}
Social network alignment, \aka anchor link prediction \cite{Liu2017User}, is to align different social networks on their common users.
To our knowledge, this problem is first proposed and addressed by the study \cite{zafarani2009connecting}.
Some studies \cite{zafarani2013connecting,kong2013inferring,liu2014hydra,mu2016user,Ren2019Meta} incorporate heterogeneous information to discover the latent consistency of user identity, while others \cite{tan2014mapping,zhang2015UMA,liu2016aligning,man2016predict,zhou2018deeplink,li2018distribution,li2017adversariallyNA} exploit the structure information to perform social network alignment. 
Moreover, some studies \cite{zhang2015cosnet,Cao2016BASS,zhang2016final,Sun2018master,Zhong2018CoLink,zhang2018mego2vec,wang2018user,xie2018unsupervised} investigate to exploit both structure and heterogeneous attribute information.
However, all existing studies consider the social network to be static for simplicity and neglect its inherent dynamics. 
The essential distinction between our work and all these studies above lies in that we for the first time pose the problem of aligning dynamic social networks and design the DNA framework to unfold the fruitful dynamics for addressing this problem.
In contrast to social network alignment, biochemical network alignment \cite{flannick2009automatic,bioinfo17dy,Bioinfo19wave} aims to find similar conserved regions between networks instead of exploring underlaying identities of nodes and thus distinguishes itself with our work.

\subsection{Network Embedding}
Network embedding is to obtain a low-dimensional compact vector representation for each node in a network.
Inspired by the success of word2vec, DeepWalk \cite{perozzi2014deepwalk} conducts random walks on the network and feeds generated node sequences into Skip-gram. 
Following DeepWalk, node2vec \cite{grover2016node2vec} proposes a novel random walk strategy to facilitate network embedding. 
LINE \cite{tang2015line} gives an explicit objective function of the first- and second-order proximity for network embedding. 
Neural networks on graph, such as GCN \cite{kipf2017semi}, also obtain the embeddings for each node.
Some studies explore the structural regularity regarding 
the regular equivalence \cite{tu2018deep}, the embedding distribution \cite{pan2018age} and the computational challenges \cite{kdd_Lian0ZGCT018},
while others attempt to incorporate the heterogeneous information  \cite{yang2015network,huang2017label}.
Recently, a few works investigate the problem of dynamic network embedding.
However, they focus on positioning the new arrivals \cite{du2018dynamic,zhang2018timers}, facilitating the embedding in one network \cite{ZhouYR0Z18,ZuoLLGHW18} or tailoring the embedding for other tasks, e.g., anomaly detection \cite{yu2018netwalk}.
The surveys \cite{cui2018survey,zhang2018network} give the comprehensive summary. 
Distinguishing with these studies, we focus on user embedding
with complex dynamics for dynamic social network alignment.


\section{Conclusion}
To the best of our knowledge, this is the first study on the problem of aligning dynamic social networks. 
We propose to reveal the evolving pattern of the individuals in the dynamic social network to facilitate aligning, inspired by the social psychology. 
The dynamics in the social network contains the discriminative pattern to facilitate alignment. However, revealing such pattern faces significant challenges in both modeling and optimization. 
Towards this end, we propose a novel DNA framework, a unified optimization approach over deep neural architectures, unfolding the fruitful dynamics for alignment.
In the DNA framework, to capture the intra-network dynamics, we propose a novel neural architecture in account of user dynamics.
To address the inter-network alignment, we design a unified optimization interplaying the embedding spaces to construct the common subspace.
To address this optimization problem, we design an effective alternating algorithm with solid theoretical guarantees. 
We conduct extensive experiments on real-world datasets and experimental results demonstrate that DNA substantially outperforms the state-of-the-art methods.



\section{Acknowledgement}
This work was supported in part by: 
National Key Research and Development Program of China under Grant 2018YFB1003804, 
National Natural Science Foundation under Grant 61602050 and 61921003, 
Fundamental Research Funds for the Central Universities 2019XD11 
and NSF under grants III-1526499, III-1763325, III-1909323, CNS-1930941, and CNS-1626432.

\bibliographystyle{IEEEtran}
\bibliography{BigData2019}

\end{document}